\documentclass[aps,prb, twocolumn,showpacs,preprintnumbers,superscriptaddress]{revtex4}

\usepackage{graphicx}
\usepackage{dcolumn}
\usepackage{bm}

\begin{document}


\title{Structural, physical and photocatalytic  properties of  mixed-valence double-perovskite   Ba$_{2}$Pr(Bi,Sb)O$_{6}$ semiconductor synthesized by citrate pyrolysis technique}

\author{Arisa Sato}
\author{Michaki Matsukawa} 
\email{matsukawa@iwate-u.ac.jp }
\author{Haruka Taniguchi}
\author{Shunsuke Tsuji}

\author{Kazume Nishidate}
\author{Sumio  Aisawa}
\affiliation{Faculty of Science and Engineering, Iwate University, Morioka 020-8551, Japan}

\author{Akiyuki Matsushita}
\author{Kun Zhang}
\affiliation{National Institute for Materials Science, Ibaraki 305-0047, Japan}

\date{\today}

\begin{abstract}
We demonstrated crystal structures, magnetic, optical, and  photocatalytic properties of the B-site substituted double perovskite Ba$_{2}$Pr(Bi$_{1-x}$Sb$_{x}$)O$_{6}$ ($x$=0, 0.1, 0.2, 0.5 and 1.0) synthesized  by the  citrate pyrolysis method.
The single-phase polycrystalline samples with  the light Sb substitution crystallized  in  a monoclinic structure ($I2/m$).
Magnetization measurements on all the samples  showed that the effective magnetic moments are concentrated  around 3  $\mu_{B}$, indicating the valence mixing states between Pr$^{3+}$ and Pr$^{4+}$. 
 The magnitudes of band gap energy for the two end member samples  were estimated from the optical measurements to be $E_{g}$ =1.06 eV at $x$=0  and  2.71 eV at $x$=1.0.  
 The Ba$_{2}$Pr(Bi$_{1-x}$Sb$_{x}$)O$_{6}$ ($x$=0, and  0.1) powders  obtained  by  the present technique  exhibited enhanced photocatalytic activities when compared to the same compounds prepared by the conventional solid state method.   Our findings suggest that the higher photocatalytic activities  strongly depend  on powder preparation methods as well as the band gaps and photogenerated charge separation. 

\end{abstract}


\renewcommand{\figurename}{Fig.}
\maketitle

\section{INTRODUCTION}
A large number of double perovskite oxides A$_{2}$B$^{'}$B$^{''}$O$_{6}$ have been widely studied due to their attractive physical properties and potential applications\cite{VA15}.
 For example, the A$_{2}$FeMoO$_{6}$ compound shows negative tunneling magnetoresistance effect at room temperature, which is of great interest with a wide range of applications in magnetic devices \cite{KO98}.
Furthermore,  multiferroic double perovskite oxides with a coupling between spontaneous ferroelectric polarization and ferromagnetic ordering are promising materials  from view points of physics and its applications. \cite{RA07}.
Some of semiconducting A$_{2}$B$^{'}$B$^{''}$O$_{6}$ compounds exhibit photocatalytic properties such as  hydrogen generation by water splitting
and are taken as alternative materials for TiO$_{2}$ oxide \cite{FU72,EN03}.
In particular, Ba$_{2}$PrBiO$_{6}$  has been shown to possess high photocatalytic activity, which is probably related to the valence mixing\cite{HA10,MA12}.   
A previous study on the magnetic states of the Ba$_{2}$PrBiO$_{6}$ compound suggests an anomalous valence situation for Pr ions.\cite{HA95} 
The complicated ground states of rare earth ions such as the  Pr ion under the crystal field effect remain open question, not only in the physical properties of the double perovskite compound, but also from the view point of physics of the $4f$ electron systems. 

There are significant factors such as charge and size differences between B$^{'}$ and B$^{''}$ sites, to determine the B-site ordering of the double perovskite oxide\cite{VA15}. 
Increase in lattice strain and/or increase in  the electrostatic repulsion overcome the entropy contribution toward disordering, causing the alternate arrangement. 
For  A$^{2+}_{2}$B$^{' 3+}$B$^{'' 5+}$O$_{6}$ composition,  the B sites tend to order  with increasing the ion size difference 
$\triangle r_{B}=r_{\mathrm{B^{'}}}-r_{\mathrm{B^{''}}}$.  If $\triangle r_{\mathrm{B}}>0.2$ \AA, it is well known that the B sites of the compounds are almost ordered alternately in the all crystallographic axes\cite{VA15}.  
In the  Ba$^{2+}_{2}$Pr$^{3+}$Bi$^{5+}$O$_{6}$ compound, the B-site ionic radius difference  $\triangle r_{\mathrm{B}}=0.23$ \AA \ meets the above condition, suggesting the B-site ordering. 
($r_{\mathrm{B^{'}}}$ (Pr$^{3+}$)=0.99 \AA \ and $r_{\mathrm{B^{''}}}$ (Bi$^{5+}$)=0.76 \AA.\cite{SH76} ) 
\\*
\ \ \  In this paper, we report x-ray diffraction measurements, magnetic susceptibilities, diffuse reflectance spectra, and 2-propanol (IPA) degradation   for the Ba$_{2}$Pr(Bi$_{1-x}$,Sb$_{x}$) O$_{6}$ compounds, to determine the crystal structures, magnetic, optical, and photocatalytic properties of B-site substituted  double perovskite oxides. 
A citrate pyrolysis technique is similar to  nitrate combustion synthesis  methods\cite{CH90} and an unique route  to prepare reactive precursor mixtures through  an ignition process of  concentrated aqueous solution including metallic ions of stoichiometric  composition .  For YBa$_{2}$Cu$_{4}$O$_{8}$ with $T_{c}$=80 K and its related cuprate oxide compounds,  it has been reported that  high-quality single-phase polycrystalline  materials  are synthesized under  ambient pressure of oxygen gas at lower annealed temperatures.\cite{KO91,HA06}  We expect  that  this procedure enables to  synthesize highly homogeneous and fine powders.  
\\*
\ \ \ As for one approach to developing efficient visible light driven photocatalysts,  it is desirable to adjust the band gaps of their semiconductors to utilize a wide range of visible light\cite{CH10,SU15}.  
According to this approach,  they are controlled by the effect of Sb substitution on the parent material. 
In addition to it,  another approach for high performance is to enhance the photogenerated charge separation  in the photocatalystic materials, to avoid charge recombination between electron and hole\cite{CH10,OH13}.  In our research, the  valence mixing states between Pr$^{3+}$ and Pr$^{4+}$ are closely related to the photo induced charge separation. 

\begin{table*}[htb]
\begin{center}
\caption{Physical and photocatalytic properties  for Ba$_{2}$Pr(Bi$_{1-x}$,Sb$_{x}$) O$_{6}$ compounds  ($x$=0, 0.1, 0.2, 0.5 and 1.0). 
 Synthetic method,  crystal symmetry, effective magnetic moment, energy gap, surface area, and gas evolution of CO$_{2}$ are listed as a function of Sb content.  The value of  evolved  CO$_{2}$ is evaluated after a visible light irradiation of 120 min.  In details, see the text. (* see  the caption of Fig.\ref{KM}($b$)) }

\begin{tabular*}{160mm}{p{25mm}cccccccccccc} \hline \hline
Sb content $x$&&Synthetic method&&Crystal symmetry &&$\mu_{\mathrm{eff}}$  && $E_{\mathrm{g}}$ && BET && CO$_{2}$   \\
	&&  && &&  ($\mu_{\mathrm{B}}$) && (eV) && (m$^{2}$/g) && (ppm/m$^{2}$)  \\ \hline
0.0 &&citrate pyrolysis && monoclinic && 3.08 &&1.06 && 1.31 &&111  \\
0.1 && citrate pyrolysis&&monoclinic && 3.05 &&1.13 && 2.88 &&102  \\
0.5 &&citrate pyrolysis&& mon./cubic && 3.07 && 1.43/1.12* &&2.72 &&$<$5   \\
1.0 &&citrate pyrolysis&& cubic &&3.0 && 2.71 && -&&$<$5   \\  \hline
0.0 && solid-state reaction&&monoclinic&& 3.15$^{a}$ &&1.1$^{a}$&& 1.23&&42   \\ 
0.1 && solid-state reaction&&monoclinic&& 3.15$^{a}$ &&1.17$^{a}$&& 1.58&& 33  \\ 
0.2 && solid-state reaction&&monoclinic&& 3.12$^{a}$ &&1.24$^{a}$&& 1.49&& 22  \\ 
1.0 && solid-state reaction&&rohmbohedral&& - &&2.22$^{a}$&& -&&$\sim$5  \\
\hline \hline
$^{a}$ see ref. \cite{TA19} \\
\end{tabular*}
\end{center}
\label{T1}
\end{table*}



\begin{figure*}[ht]
\begin{center}
    \begin{tabular}{c}
      \begin{minipage}[t]{0.5\hsize}
      \begin{center}
        \includegraphics[width=9cm, pagebox=cropbox, clip]{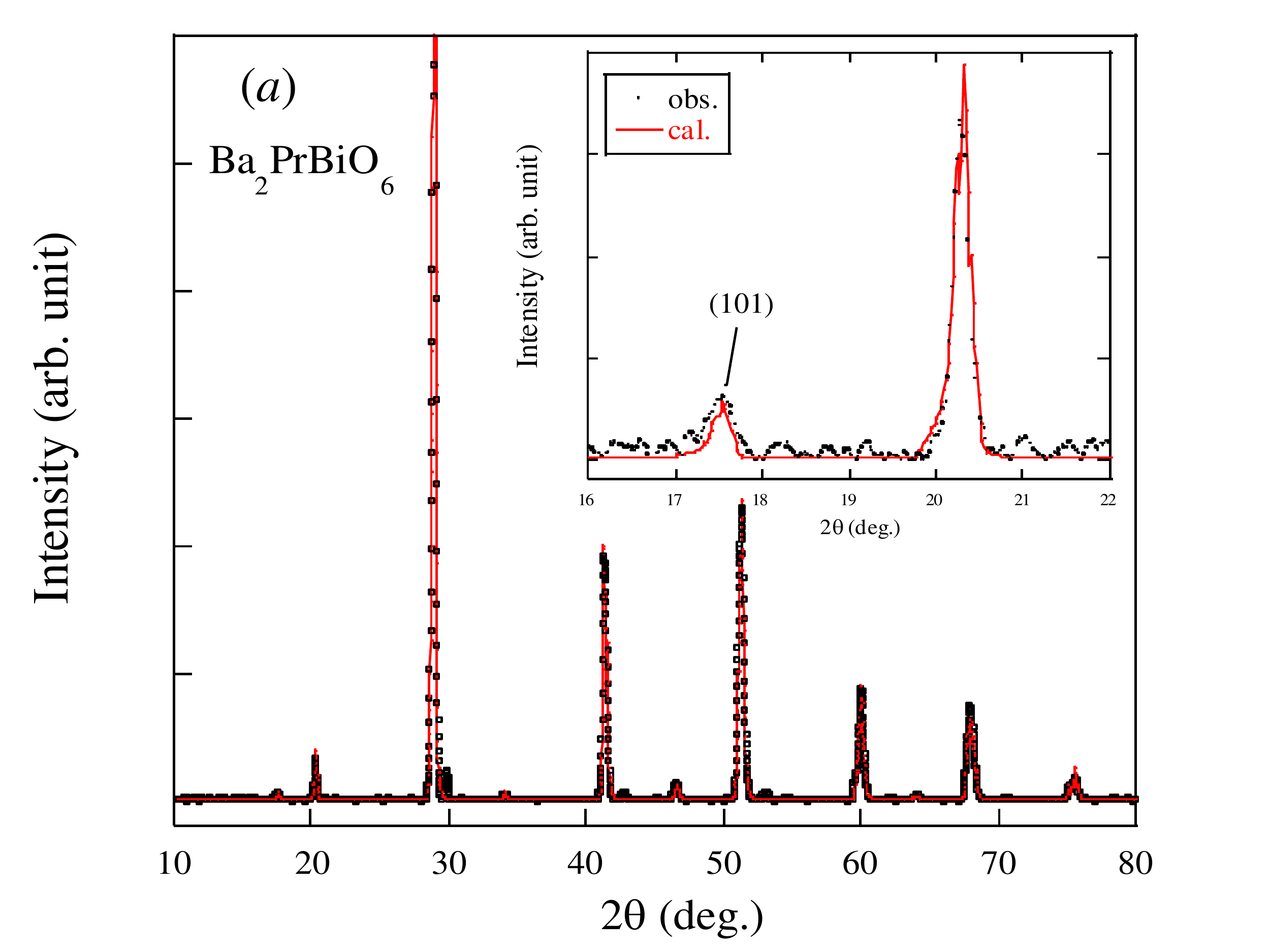}
 \end{center}  
      \end{minipage} 
      \begin{minipage}[t]{0.5\hsize}
      \begin{center}
        \includegraphics[width=9cm, pagebox=cropbox, clip]{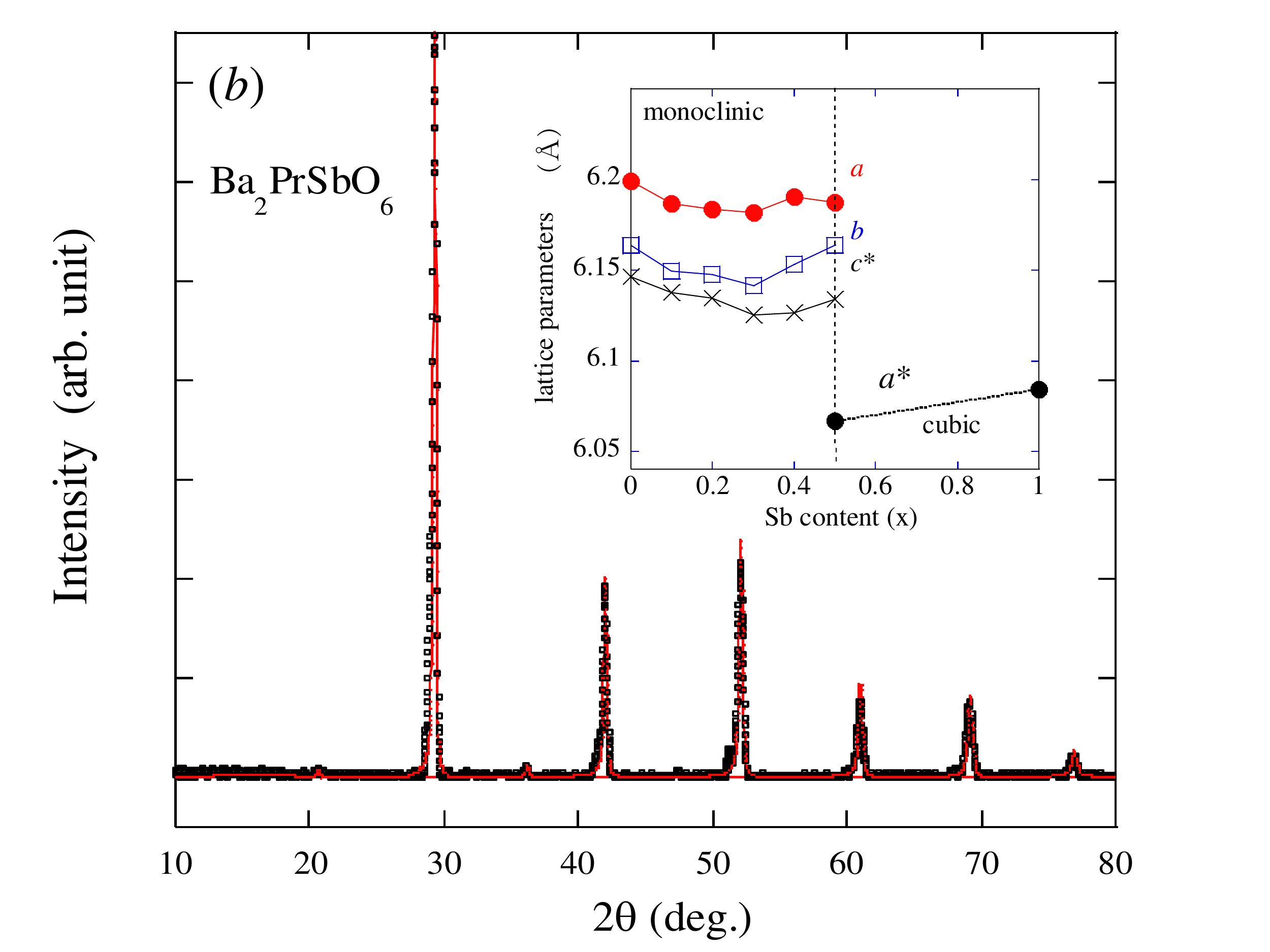}
 \end{center}  
      \end{minipage}
    \end{tabular}
     \begin{tabular}{c}
      \begin{minipage}[t]{0.5\hsize}
      \begin{center}
        \includegraphics[width=9cm, pagebox=cropbox, clip]{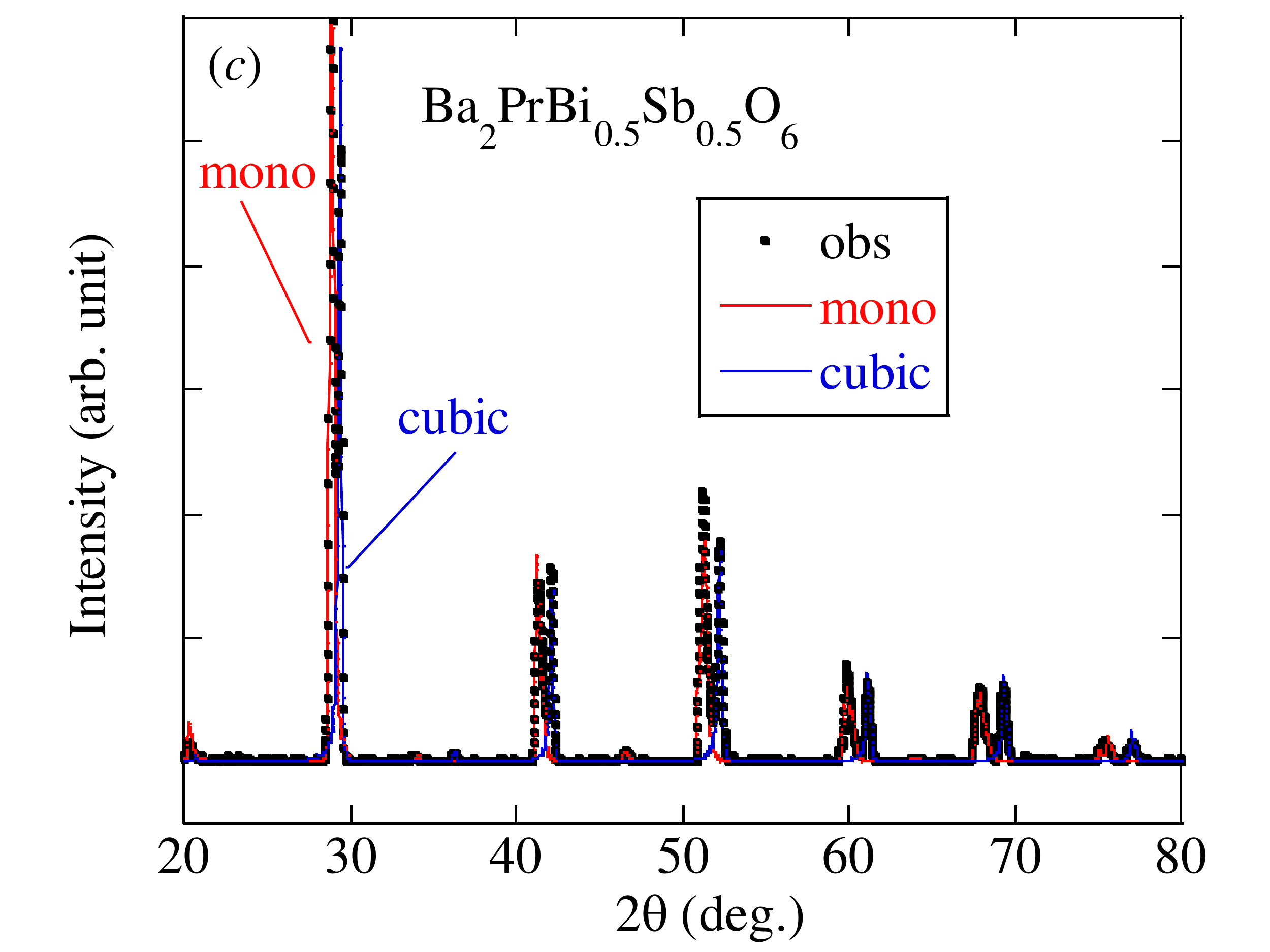}
 \end{center}  
      \end{minipage} 
      \begin{minipage}[t]{0.5\hsize}
      \begin{center}
        \includegraphics[width=9cm, pagebox=cropbox, clip]{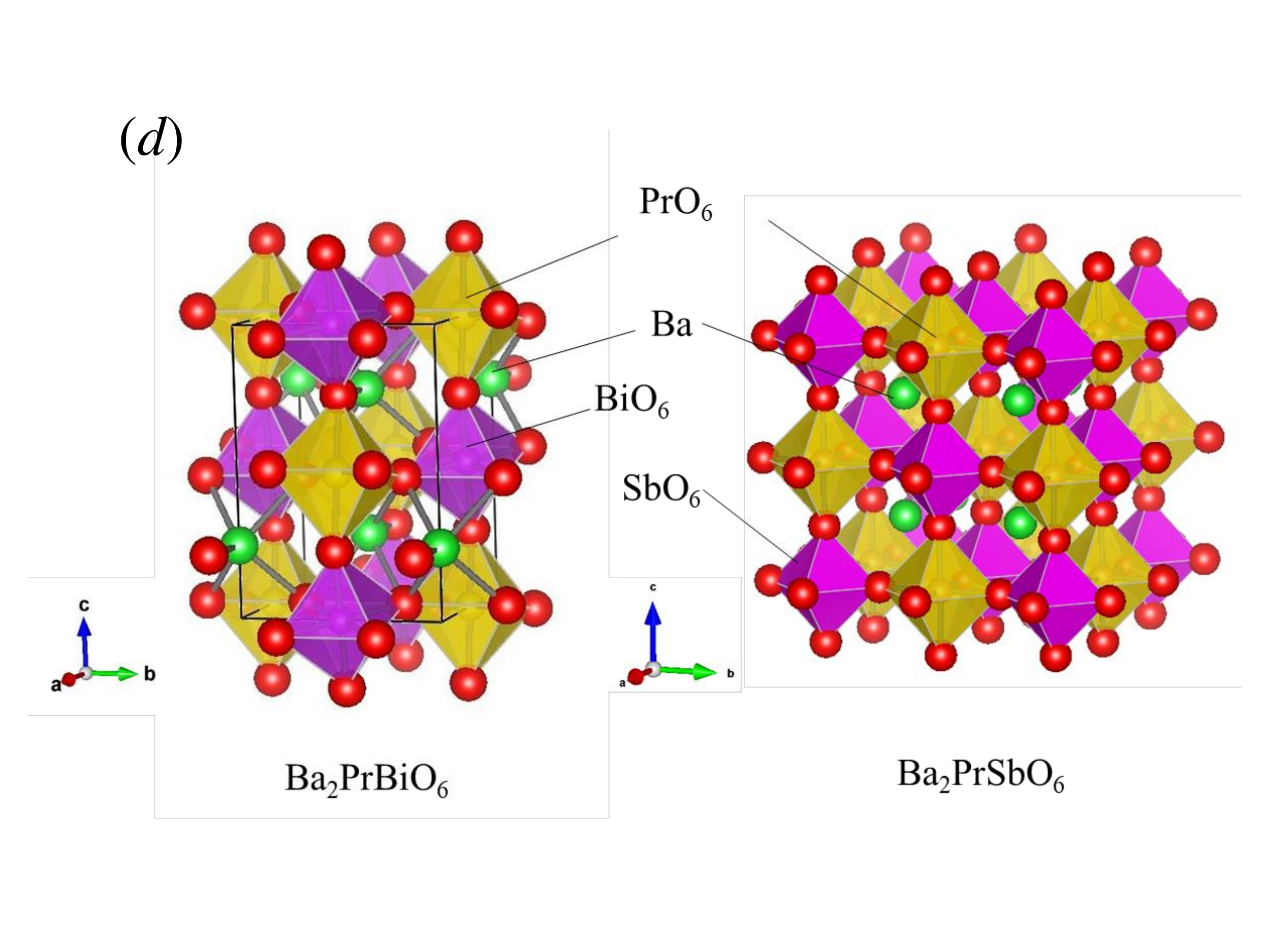}
 \end{center}  
      \end{minipage}
    \end{tabular}
\caption{(color online) X-ray diffraction patterns for the two end-member citrate samples,  ($a$) Ba$_{2}$PrBiO$_{6}$  and ($b$)  Ba$_{2}$PrSbO$_{6}$.  In addition,  the data for the intermediate sample Ba$_{2}$Pr(Bi$_{0.5}$,Sb$_{0.5}$)O$_{6}$ are given in ($c$). The calculated profiles based on the monoclinic and cubic structure models  are shown in ($a$) and ($b$), respectively. The insets of ($a$) and  ($b$) display the enlarged diffraction profiles and the lattice parameters as a function of Sb content. In the inset of  ($b$), for monoclinic and cubic structures,  $c*=c/\sqrt{2}$ and $a*=a/\sqrt{2}$, respectively. In  ($c$).  the red and blue plots are responsible for  the monoclinic and cubic structure models, respectively. ($d$) Crystal structures  for the two end-member samples,   Ba$_{2}$PrBiO$_{6}$  and Ba$_{2}$PrSbO$_{6}$.  The left and right pictures represent the ideal monoclinic and cubic structures, respectively.  Bi(or Sb)O$_{6}$ and  PrO$_{6}$ octahedra are connected by corner-sharing with each other. }
 \end{center}  
\label{Xray} 
\end{figure*}

\begin{figure}[ht]
\includegraphics[width=9cm, pagebox=cropbox, clip]{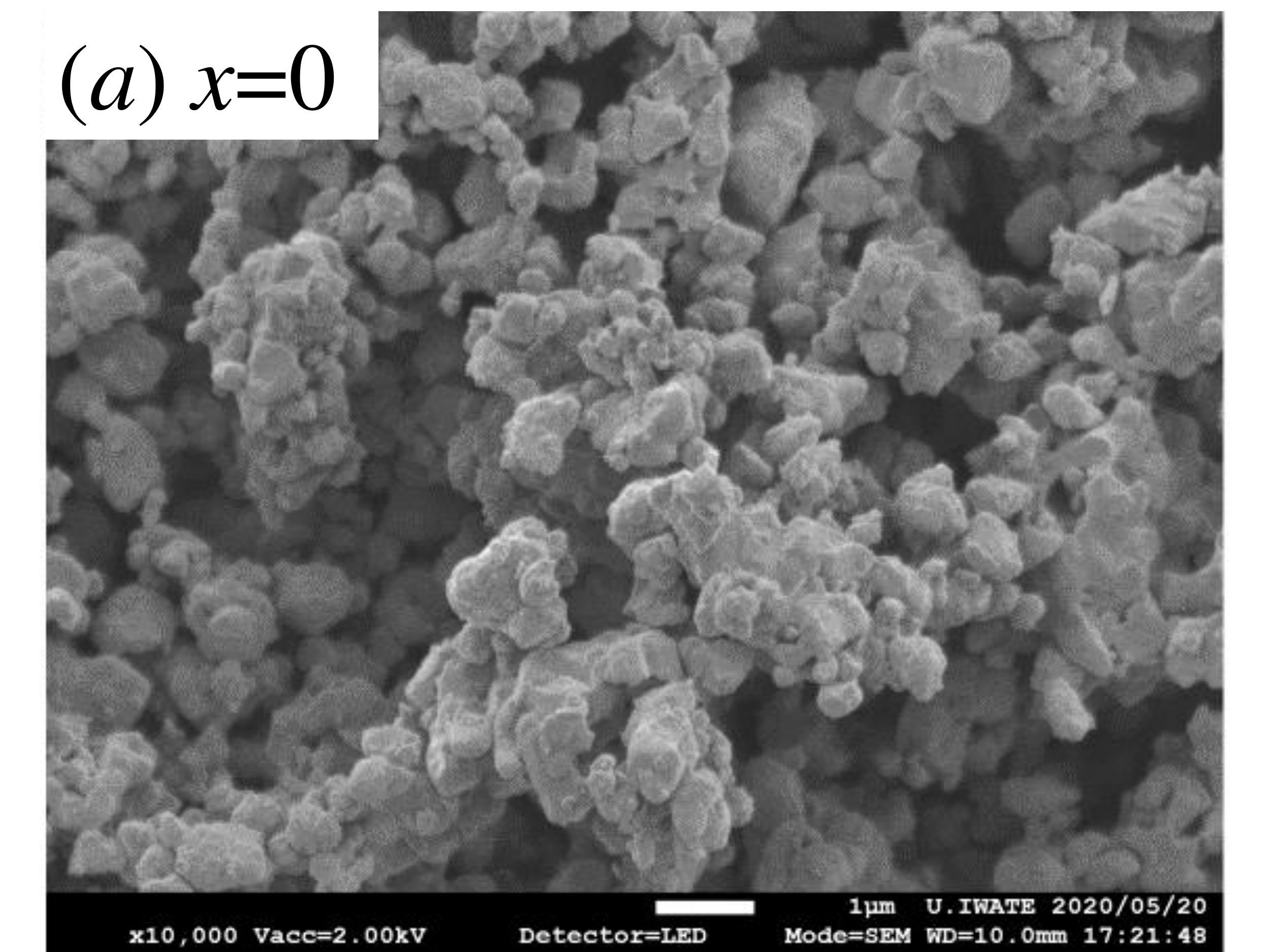}
\includegraphics[width=9cm, pagebox=cropbox, clip]{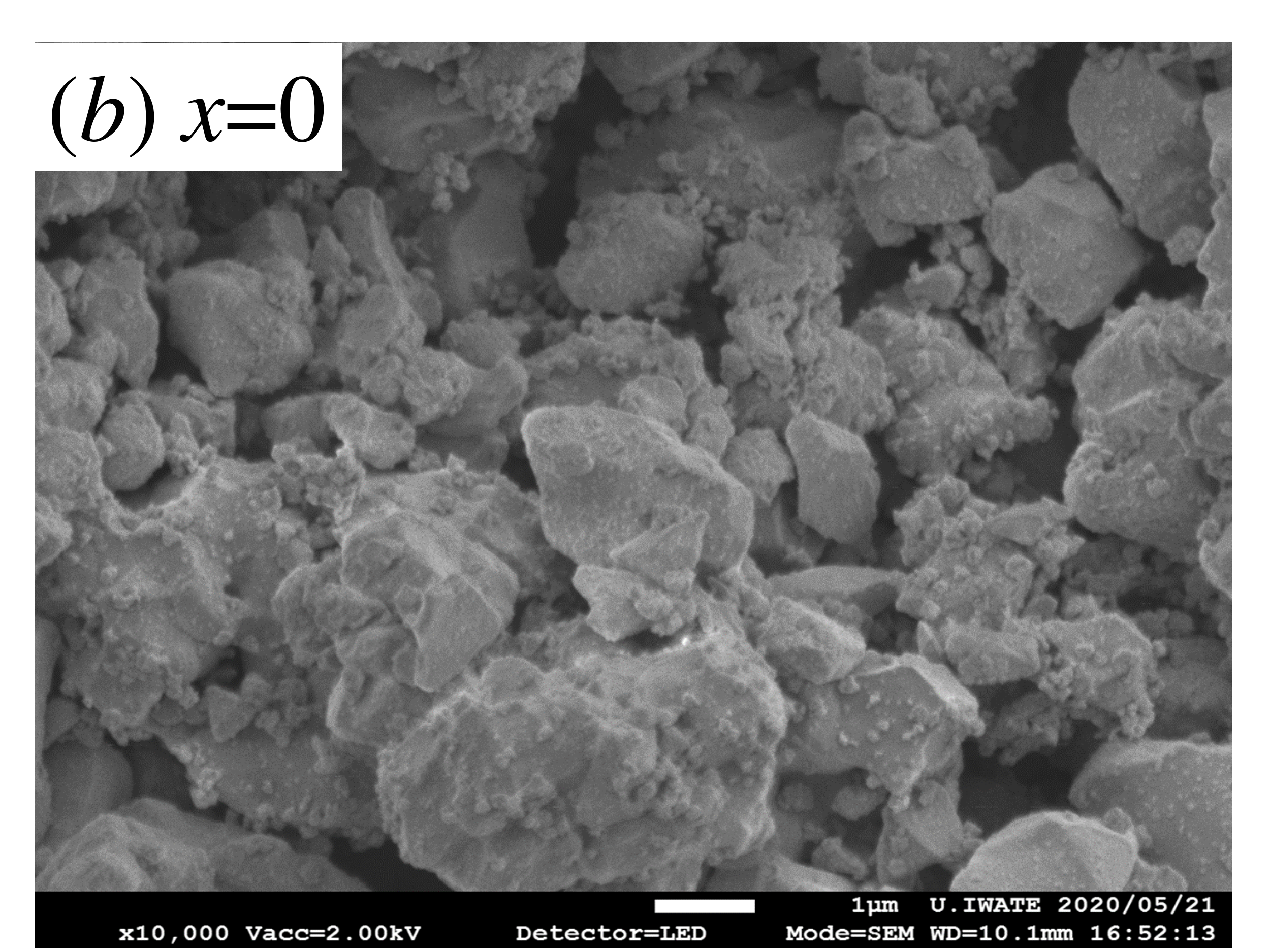}
\caption{(color online)  SEM images of   Ba$_{2}$PrBiO$_{6}$ powders.  ($a$) citrate pyrolysis  and  ($b$) solid-state reaction techniques.  (For Ba$_{2}$Pr(Bi$_{1-x}$,Sb$_{x}$)O$_{6}$ with $x$= 0.1, and 1.0, see supplementary data Fig.S2) }
\label{SEM}
\end{figure}

\begin{figure}[ht]
\centering
\includegraphics[width=9cm, pagebox=cropbox, clip]{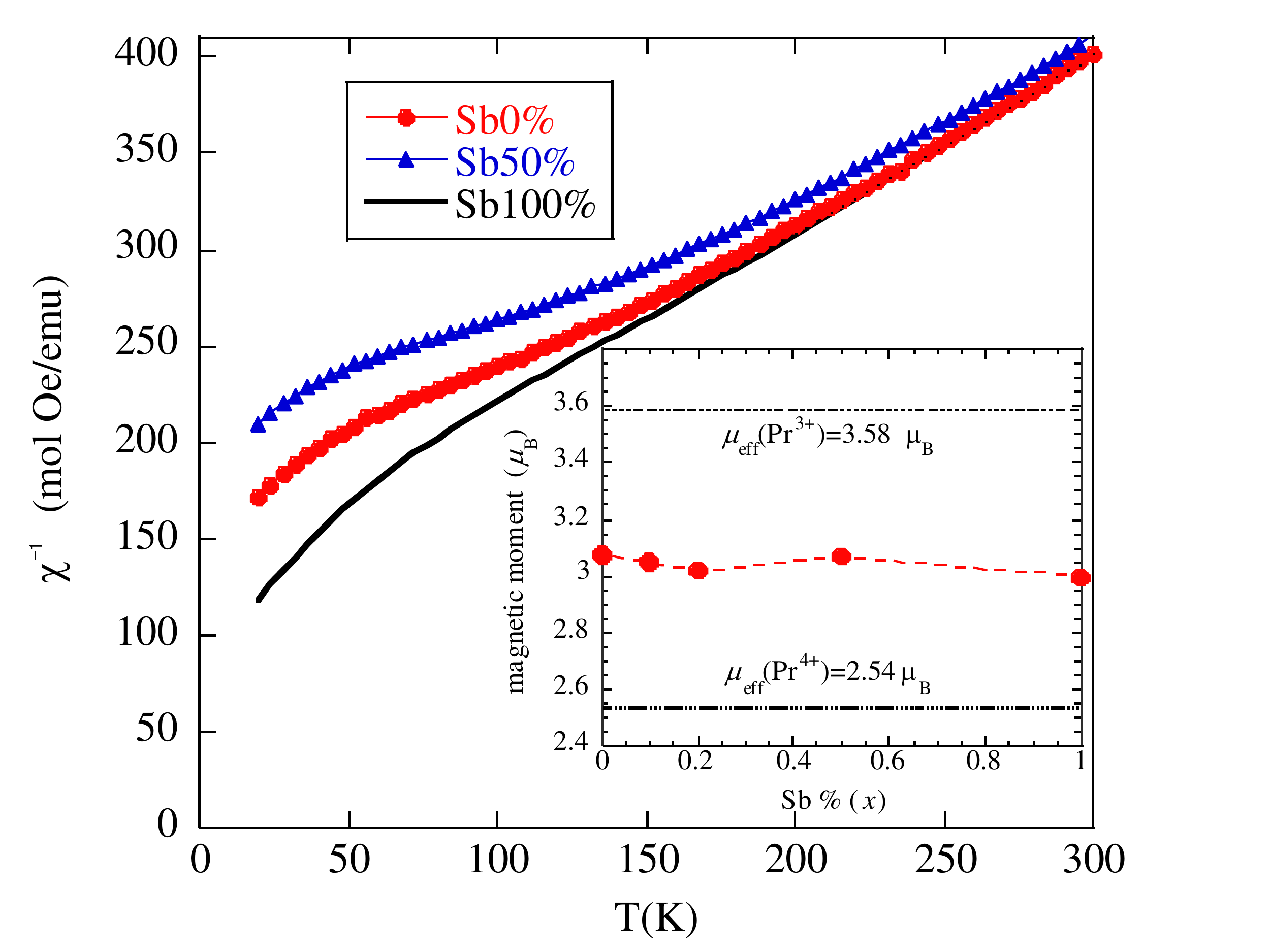}
\caption{(color online) Inverse magnetic susceptibility vs temperature. ($\chi^{-1} $) for  Ba$_{2}$Pr(Bi$_{1-x}$,Sb$_{x}$) O$_{6}$($x$=0, 0.5, and 1.0).  The inset denotes the effective magnetic moment plotted as a function of Sb content. The value of $\mu_{\mathrm{eff}}$ is evaluated  using the Curie-Weiss law. The  lower and upper  limits represent  $\mu_{\mathrm{eff}}$(Pr$^{4+}$)=2.54 $\mu_{B}$ and  $\mu_{\mathrm{eff}}$(Pr$^{3+}$)=3.58 $\mu_{B}$, respectively.
}
\label{MT}
\end{figure}

\section{EXPERIMENT}
Polycrystalline samples of Sb-substituted Ba$_{2}$PrBiO$_{6}$  were synthesized  by using  the citrate pyrolysis technique. In the first step,  stoichiometric mixtures of high purity BaCO$_{3}$, Pr$_{6}$O$_{11}$, Bi$_{2}$O$_{3}$, and Sb  were dissolved in a nitric acid solution at 70-80 $^{\circ} $C. After adding  citric acid and neutralizing it by aqueous ammonia,  we then obtained  the porous products  through the self-ignition process using halogen lamp stirrer.  In the next step,   the  precursors  were  ground and resultant fine powders were  annealed  in air at 900-1000 $^{\circ} $C for 48-96h, in order to synthesize the Ba$_{2}$PrBiO$_{6}$ double perovskite  phase.  
For scanning electron microscope measurements,   the  Ba$_{2}$PrBiO$_{6}$ polycrystalline film on Ag substrate was fabricated from the single-phase powders by an  electrophoretic deposition technique. 
The  electrophoretic deposition was conducted in the acetone and iodine bath under the application of  electric voltage up to 300 V for 120 s \cite{KA01}.  We set  Pt and Ag plates as anode and cathode electrodes, respectively.
We performed X-ray diffraction measurements on the produced samples
at room temperature with an Ultima IV diffractometer (Rigaku) using Cu-K$\alpha $ radiation. 
The lattice parameters were estimated  from the x-ray diffraction data using the RIETAN-FP program.
The Brunauer-Emmett-Teller (BET) surface area of the powder samples was evaluated  using a surface area analyzer (BELSORP-mini I\hspace{-.1em}I , Microtrac). 

Optical spectra  were measured by a diffuse reflectance method using a spectrophotometer (Hitachi U-3500) and BaSO$_{4}$ was used as the reference material.  The optical band gaps for the powder samples were evaluated from reflectance spectral data using the conventional Kubelka-Munk functions\cite{HA10,TA07} .
The $dc$ magnetization was measured  at a magnetic field of 0.1 T under the field cooling process using a superconducting quantum  interference device  magnetometer (MPMS,Quantum Design). 
We conducted the gaseous 2-propanol (IPA)  degradation experiment, to evaluate photocatalytic activities  of  the  powder samples. (in detail, refer to \cite{HA10,MU04}) .  
The powder (about 1g) was placed on the bottom of a small glass cell and its cell was set in a 0.5-L glass reactor vessel. 
After removing impurities from the vessel through filling a dry air,  the  dilute IPA gas (5cc) was injected into its vessel  using a syringe. 
We started illuminating the visible light after the IPA  concentration remained constant, which implied that the IPA gas finished absorbing on the surface of particles. 
The visible light   was obtained  using  a 300W  Xe lamp equipped with UV and IR filtering functions (Cermax LX300F, Excelitas Technologies). The illuminating spectra of the Xe lamp are limited in  the visible wavelength range between 390 and 780 nm (see supplementary data Fig. S1). 
 It is well known that the IPA gas under photocatalytic reaction is finally decomposed into  CO$_{2}$  \cite{MU04}.  Accordingly,  the  CO$_{2}$ concentration was measured as a function of  irradiation time using a gas chromatography system  (GC-2014, Shimazu Co.)     

\section{RESULTS AND DISCUSSION}
The X-ray diffraction patterns for  the two end-member oxides, Ba$_{2}$PrBiO$_{6}$  and  Ba$_{2}$PrSbO$_{6}$, are shown in Fig.\ref{Xray}\ ($a$) and ($b$), respectively. For the parent Ba$_{2}$PrBiO$_{6}$ with  a monoclinic structure (the space group $I2/m$),  the lattice parameters are estimated to be 
 $a=6.2038$ \AA , $b=6.1689$ \AA, $c=8.7011$ \AA \ and $\beta =89.7303 ^{\circ}$  from the x-ray diffraction data using RIETAN-FP program, which are in good agreement with previous data\cite{HA95, ON18}. The emergence of (101) reflection (the inset of Fig.\ref{Xray}($a$)) indicates B-cation ordering which is characteristic of the ordered double-perovskite structure. 
In particular, the tiny profiles around the  (101) peak  are well described by the least squared method based on RIETAN-FP when we set B site ordering to be  $\sim$70 $\%$.  This estimated ratio is nearly in good agreement with the published value ( $\sim$75 $\%$)\cite{HA95} . 
The polycrystalline samples for $x<0.5$  are formed in almost single phases of the monoclinic structure, while  the $x=1.0$ sample crystallizes in a cubic structure with the space group 
$Fm\overline{3}m$.  Substitution of the smaller Sb$^{5+}$ (0.60 \AA) ion at the Bi$^{5+}$ (0.76 \AA) site causes  a monotonic decrease in the lattice parameters as displayed in the inset of Fig. \ref{Xray} ($b$). 
Upon further increasing  Sb content,  the x-ray diffraction patterns for  $x$=0.5 sample exhibit a two-phase mixture of  monoclinic and cubic structures (Fig.\ref{Xray}($c$)). 
The  double main peaks responsible for the monoclinic and cubic phases
are almost comparable and they are in the  ratio of   56 $\%$ ($I2/m$) to  44  $\%$ ($Fm\overline{3}m$) . 
For Ba$_{2}$PrSbO$_{6}$, we obtain the lattice parameters  $a=8.6042$ \AA \ and $\alpha = 90^{\circ}$. Our result is qualitatively different from  the rhombohedoral structure as previously reported  \cite{FU05,OT15}.   The differences in crystal structures are attributed to the tolerance factors which are associated with the mixed valence state of Pr ions.   We will discuss this interesting issue following the magnetic data. 
The  crystal structures  for the two end-member samples,   Ba$_{2}$PrBiO$_{6}$  and Ba$_{2}$PrSbO$_{6}$ are shown in Fig.\ref{Xray}($d$),  corresponding to  the monoclinic and cubic structures, respectively.  The Bi(or Sb)O$_{6}$ and  PrO$_{6}$ octahedra are connected by corner-sharing with each other. 
 However, the corresponding  (101) peak observed at the low Sb substituted  sample is strongly suppressed at $x$=1.0, indicating the B-site disordering. 
  
The microstructures  of the parent samples prepared by the citrate pyrolysis and solid-state reaction methods are shown in Fig.\ref{SEM}. 
 (For the Sb substituted samples with $x$= 0.1 and 1.0, see supplementary data Fig.S2)
The particles of the former sample have an average size in the range of  about 0.2-0.5 micron.  On the other hand,  the particle diameters of the latter sample  are on a micron order scale and about one-order grater than those of the former.  The citrate pyrolysis process provides homogeneously dispersed grains with sub micron size in comparison to the conventional solid-state procedure. 
We suspect that the substitution of Sb ion at Bi sites promotes the refinement of powder samples as
shown  in Fig. S2.  This finding is probably related to the larger surface area for the Sb substituted samples. 

The inverse magnetic susceptibility data  ($\chi^{-1} $) for the  Ba$_{2}$Pr(Bi$_{1-x}$,Sb$_{x}$) O$_{6}$ compounds  ($x$=0,  0.5 and 1.0) are shown in Fig.\ref{MT} as a function of temperature at a magnetic field of 0.1 T.  All the measured samples  exhibit  no signature of magnetic ordering  over a wide range of temperatures.  
From the magnetization data at high temperatures above 200 K, we estimate the effective magnetic moment according to the Curie-Weiss law, $\chi=C/(T-\Theta)$.  Here,  $C$ and $\Theta$ are the Curie constant and  Curie temperature. 
The value of effective magnetic moment $\mu_{\mathrm{eff}}$ is evaluated by using the following formula,
$$
C=N\mu_{\mathrm{eff}}^{\ \ 2}\mu_{B}^{\ 2}/3k_{B}
$$
where $N$ and $\mu_{B}$ denote the number of magnetic atom per mol and the Bohr magneton, respectively.
Performing the calculation for the parent and lightly Sb substituted samples, we obtain that $\mu_{\mathrm{eff}}$ = 3.07 $\mu_{B}$ and 3.05 $\mu_{B}$ in $x$=0 and $x$=0.1 cases, respectively.
As shown in the inset of Fig. \ref{MT},  the magnetic moment is almost stable against the Sb substitution.
Next, we try to estimate the ratio of the Pr$^{3+}$ and Pr$^{4+}$ ions  using the equation 
$$
\mu^{2}_{\mathrm{eff}}=y \mu^{2}_{\mathrm{eff}}(\textrm{Pr}^{3+})+(1-y) \mu^{2}_{\mathrm{eff}}(\textrm{Pr}^{4+})
$$
where $\mu_{\mathrm{eff}}$(Pr$^{3+}$)=3.58 $\mu_{B}$ and $\mu_{\mathrm{eff}}$(Pr$^{4+}$)=2.54 $\mu_{B}$. For the parent sample, we obtain that the ratio of Pr$^{3+}$  and Pr$^{4+}$ ions is 0.47 : 0.53. 
For $x$=0.1, Pr$^{3+}$ : Pr$^{4+}$ = 0.45 : 0.55. For all the samples,  the magnetic data strongly suggest the coexistence between trivalent and tetravalent states of Pr ions. 
The X-ray photoemission spectroscopy analysis of the solid-state parent sample \cite{ON18} revealed that a prominent peak of Pr$^{3+}$ is visible  accompanied by  a smaller shoulder structure of Pr$^{4+}$, which is consistent with the mixed valence state of the Pr ion.

The magnetic data suggest that about half of Pr ions are oxidized to the tetravalent state over the whole 
range of Sb substitution.  
Accordingly, for  the  end members  with $x$ =0 and 1.0,  we assume that  Ba$^{2+}_{2}$Pr$^{4+}$M$^{3+}_{0.5}$M$^{5+}_{0.5}$O$_{6}$ (M= Bi or Sb) is probably realized  under the charge neutrality in contrast to  Ba$^{2+}_{2}$Pr$^{3+}$M$^{5+}$O$_{6}$.  
First of all,  we consider B site disordering of the parent sample on the basis of  the B-site ionic radius difference  $\triangle r_{B}$ as mentioned in the part of Introduction. 
For the  Ba$^{2+}_{2}$Pr$^{4+}$Bi$^{3+}_{0.5}$Bi$^{5+}_{0.5}$O$_{6}$ composition,  $\triangle r_{B}$  was decreased from 0.23 \AA \ at the trivalent state of Pr ion  to 0.045 \AA \ at the tetravalent state,
where  $r_{\mathrm{B^{'}}}$ (Pr$^{4+}$)=0.85 \AA \ and $r_{\mathrm{B^{''}}}$ (Bi$^{3+}_{0.5}$Bi$^{5+}_{0.5}$)=0.895 \AA \cite{SH76}. 
In addition to the smaller ion radius difference,   the difference in the $B^{'}$ and $B^{''}$  cation oxidation states  $\triangle Z_{B}$  becomes  zero  in the case of Pr$^{4+}$.  For Pr$^{3+}$ and Bi$^{5+}$, $\triangle Z_{B} = 2$.
Considering the electrostatic repulsion between B site cations,  when  $\triangle Z_{B} $ is large, the ordered arrangement of B site cations gives lowered energy compared to the disordered ones \cite{VA15}. 
From the view point of  these two  factors,  we expect that  the  Pr$^{4+}$ based phase  favors  the B-site disordering, which is consistent with the small (101) peak in the X-ray diffraction data in Fig. \ref{Xray}(a). 
In the parent compound,  the mixed valence state of Pr ions is closely related to the partial  disordering of B-site. 

Next,   we discuss the tolerance factors of the two end member samples  because they  are associated with the crystal structures  through  the ionic radii of Pr, Bi and Sb. 
The tolerance factor of double perovskite compounds Ba$_{2}$PrMO$_{6}$  is given by the following equation,
$$
t=\frac{r_{\mathrm{Ba}}+r_{\mathrm{O}}}{\sqrt{2}\left(\frac{r_{\mathrm{Pr}}+r_{\mathrm{M}}       }{2}+r_{\mathrm{O}}\right)}
$$
where $r_{\mathrm{Ba}}$, $r_{\mathrm{O}}$, $r_{\mathrm{Pr}}$, and $r_{\mathrm{M}}$ are the ionic radii of the respective ions. 
For (Pr$^{3+}$,\ M$^{5+}$\ =\ Bi$^{5+}$) and (Pr$^{4+}$,\ M$^{4+}$\ =\ Bi$^{3+}_{0.5}$Bi$^{5+}_{0.5}$),  we obtain $t=0.9356$ and 0.9366, respectively.  These values remain below  the threshold value $t_{\mathrm{mono}}=\sim 0.96$ for  the phase boundary between  monoclinic and rhombohedral structures in the phase diagram of double perovskite compounds Ba$_{2}$PrMO$_{6}$ \cite{OT15}. 
On the other hand,  for  (Pr$^{3+}$,\ M$^{5+}$\ =\ Sb$^{5+}$) and (Pr$^{4+}$,\ M$^{4+}$\ =\ Sb$^{3+}_{0.5}$Sb$^{5+}_{0.5}$),  we have $t=0.9697$ and 0.9831, respectively.   It is true that the former value is below the threshold value $t_{\mathrm{rhom}}=\sim 0.974$ for  the phase boundary between rhombohedral and cubic structures, but the latter increases  above  this critical value.   This estimation strongly supports the stability of cubic structure in the Pr$^{4+}$ dominant  Ba$_{2}$PrSbO$_{6}$ compound.  Our magnetic data for the end member  Ba$_{2}$PrSbO$_{6}$ showed that  Pr$^{4+}$ ions occupy the majority of 60\%  accompanied by the minority (40\%) of Pr$^{3+}$. Furthermore,  the x-ray data for this compound also revealed  the presence of the cubic crystal structure.  
Our experimental data support  the close relationship between the valence state of  Pr$^{4+}$ and the occurrence of cubic phase through the tolerance factor. 
For the intermediate composition Ba$_{2}$PrBi$_{0.5}$Sb$_{0.5}$O$_{6}$,  we identified the monoclinic and cubic phases with Ba$_{2}$PrBiO$_{6}$ and Ba$_{2}$PrSbO$_{6}$, respectively.  
The coexistence of these two phases is also explained on the basis of the above scenario concerning the two end member compositions. 

\begin{figure}[ht]
\centering
\includegraphics[width=9cm, pagebox=cropbox, clip]{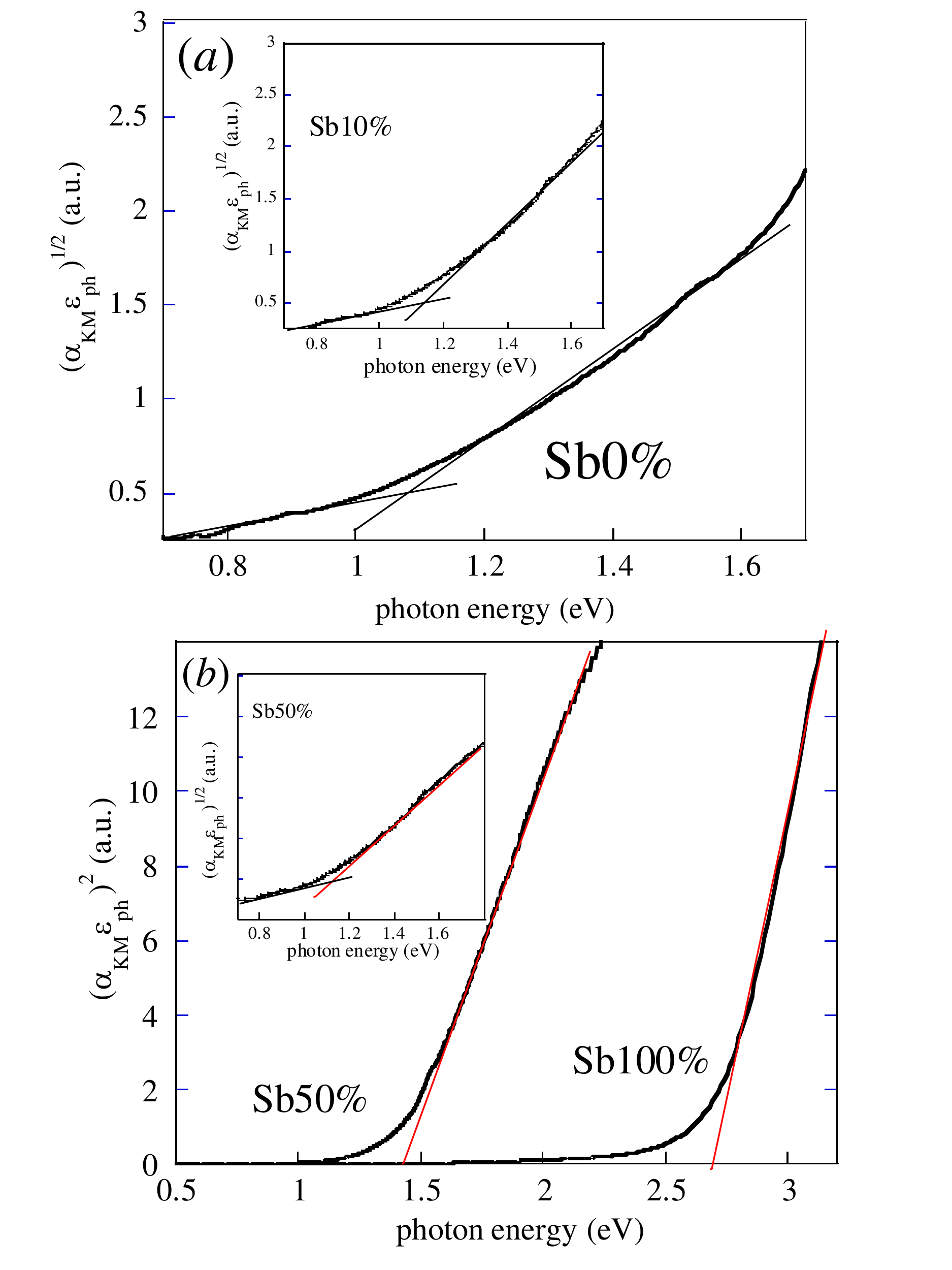}
\caption{(color online) Optical properties of  Ba$_{2}$Pr(Bi$_{1-x}$,Sb$_{x}$) O$_{6}$($x$=0, 0.1, 0.5, and 1.0).
 ($a$) Plots of $(\alpha _{\mathrm{KM}}\varepsilon _{p})^{1/2}$ vs $\varepsilon _{p}$. For the $x$=0 and 0.1 samples, $Eg$  is estimated from  an intersection point of  base line and straight line by extrapolation.
 In the inset,  the $x=0.1$ data are shown. 
  ($b$)  $(\alpha _{\mathrm{KM}}\varepsilon _{p})^{2}$vs $\varepsilon _{p}$ for the $x=0.5$ and $x=1.0$ samples  are plotted as a function of $\varepsilon _{p}$. The inset denotes plots of $(\alpha _{\mathrm{KM}}\varepsilon _{p})^{1/2}$ vs $\varepsilon _{p}$ for $x=0.5$.  The  corresponding optical data  are  fitted using both indirect and direct transition models in the photon energy region between 1 and 2 eV. 
The former and latter band gap energies  are  estimated to be  $\sim$1.12 eV and 1.43 eV, respectively. 
}
\label{KM}
\end{figure}

\begin{figure}[ht]
\centering
\includegraphics[width=9cm, pagebox=cropbox, clip]{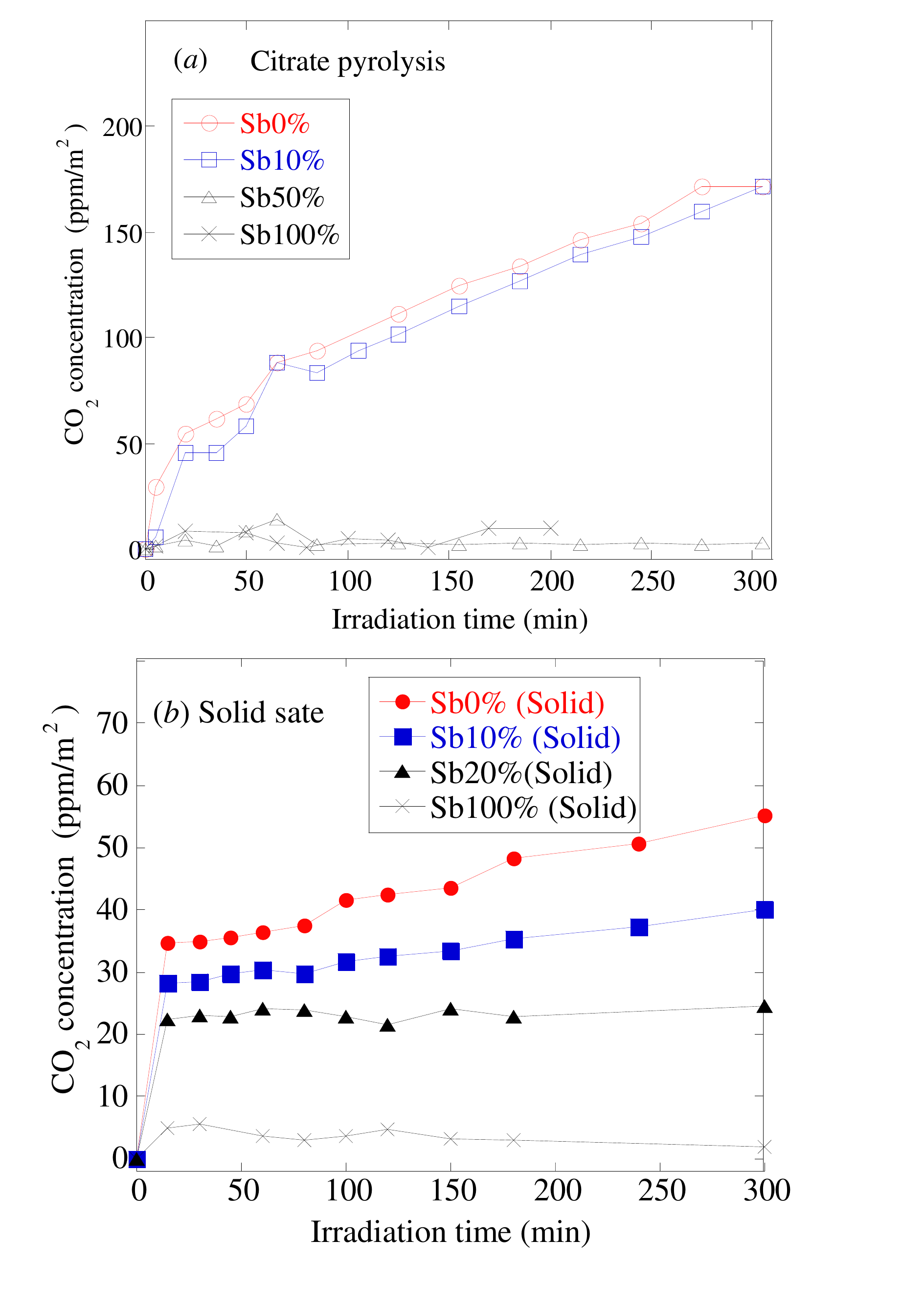}
\caption{(color online)Photocatalytic activities vs visible light irradiation time for  Sb-substituted Ba$_{2}$PrBiO$_{6}$.  ($a$) citrate pyrolysis samples, and   ($b$) solid-state samples.   The gaseous concentration  is normalized by the surface area of the powder samples listed in Table \ref{T1}.   }
\label{CO2}
\end{figure}

\begin{figure}[ht]
\centering
\includegraphics[width=9cm, pagebox=cropbox, clip]{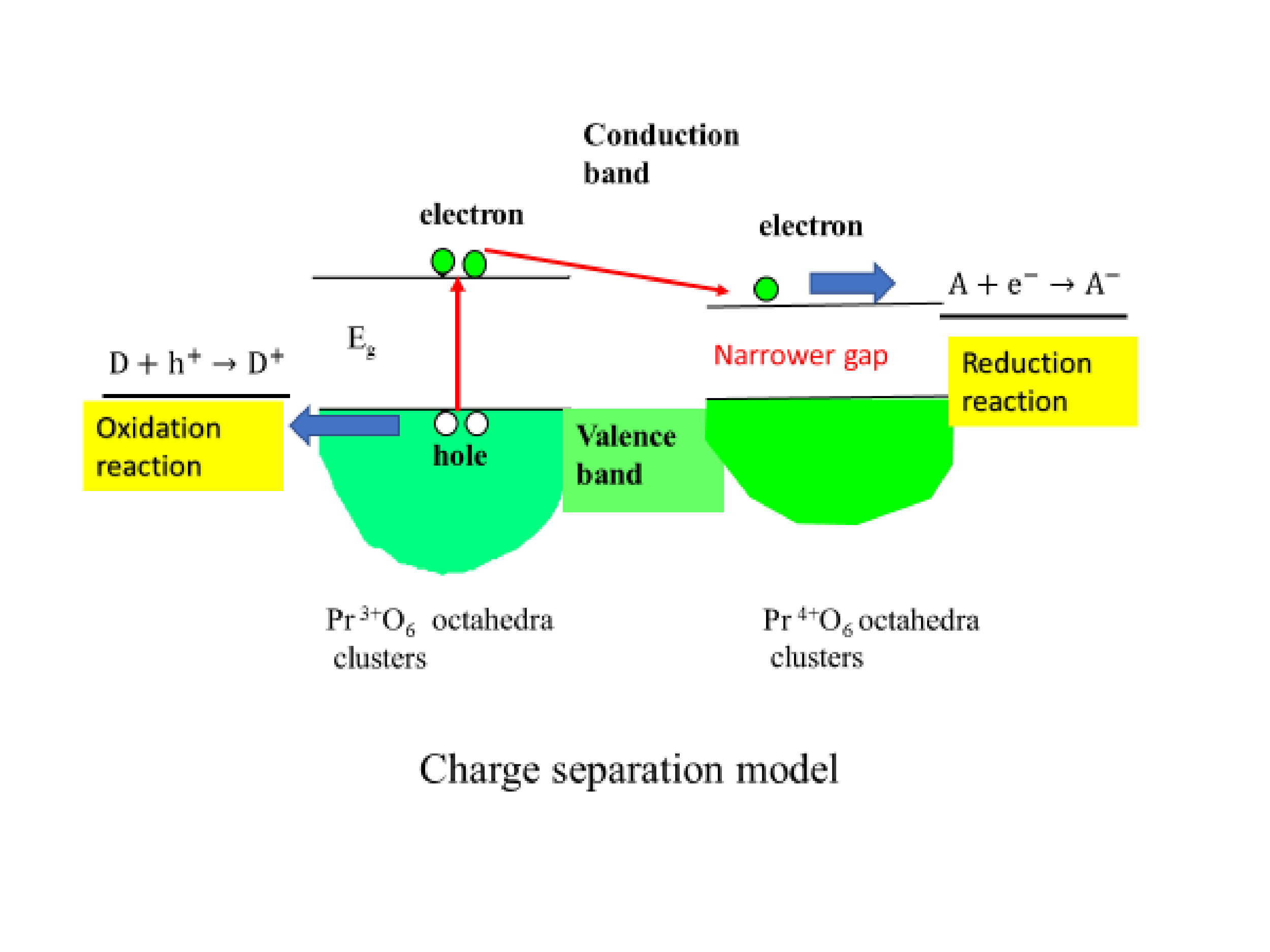}
\caption{(color online) Charge separation model for  the coexistence between trivalent and tetravalent states of Pr ions.  The band gap at the cluster region occupied by the Pr$^{4+}$O$_{6}$ octahedra  is  assumed to be narrower than that at its neighboring region occupied by the Pr$^{3+}$O$_{6}$ octahedra. (see the text)  }
\label{Model}
\end{figure}



We demonstrate the optical properties
 for the Ba$_{2}$Pr(Bi$_{1-x}$,Sb$_{x}$)O$_{6}$ ($x$=0, 0.1,  0.5, and 1.0) powder samples measured by  the diffuse reflectance method.
First, the observed reflectance data for the powder samples are transformed to the absorption coefficient  $\alpha _{\mathrm{KM}}$ by using the conventional Kubelka-Munk function.
Next, for the Kubelka-Munk conversion data near the band edge,  we extrapolate the tangent line to the $\varepsilon _{p}$ axis and 
evaluate the optical band gaps from the intersection according to the equation of 
$$
(\alpha _{\mathrm{}}\varepsilon _{p})^{n}\propto (\varepsilon _{p}-E_{g})
$$
, where $\alpha _{\mathrm{}}$, $\varepsilon _{p}$, and $E_{g}$ are the absorption coefficient, the photon energy, and  the band gap energy \cite{HA10,MA12}. 
Here, it should be noted that  the power exponent $n$ on the left hand side of the above formula decides the types of photon transition of optical absorption. 
For direct and indirect semiconductors, we assume that  $n=2$ and $n=1/2$, respectively \cite{HA10,MA12}. The optical measurements determine whether the energy gap is direct or indirect.
Figure \ref{KM} shows  the absorption coefficient  as a function of photon energy. For the $x$=0 and 0.1  samples, we adopt  $(\alpha _{\mathrm{KM}}\varepsilon _{p})^{1/2}$ vs $\varepsilon _{p}$ plot.

In Fig.\ref{KM} ($b$) , the square of  the absorption coefficient for the $x$=0.5 and 1.0 samples, $(\alpha _{\mathrm{KM}}\varepsilon _{p})^{2}$, is plotted as a function of $\varepsilon _{p}$. 
We estimate $E_{g}$ = 1.06 eV at $x$=0  and  2.71 eV at $x$=1.0, assuming indirect and direct photon transitions, respectively.  
 The magnitude of the energy band gap is substantially enhanced with increasing Sb content. When the Sb content exceeds 50 $\%$, the nature of photon absorption is changed from the indirect to direct transitions.  
We give some comments on the optical property  for the Sb50$\%$ substituted sample. 
As discussed above,  the crystal structure of the present sample is explained by a two-phase mixture of  monoclinic and cubic structures.   The corresponding optical data are  fitted using both indirect and direct transition models in the photon energy region between 1 and 2 eV. 
The former energy gap is estimated to be  $\sim$1.12 eV, which is similar to the values of the monoclinic single  phase $x$=0 and 0.1  samples. However, the latter (1.43 eV)  is quite  different from 2.71 eV at the cubic single phase  $x$=1.0 sample.  This reason has not  been made clear but it has some relationship with chemical  interaction between the two-phases such as partial substitution  of the Bi ions at Sb sites of the cubic phase.

Next, we conducted the gaseous IPA degradation experiment under a visible light irradiation, to evaluate photocatalytic activities  of  the  powder samples.  Figure  \ref{CO2} ($a$) shows the concentration of evolved CO$_{2}$ as a function of  visible light irradiation time for  the Sb-substituted Ba$_{2}$PrBiO$_{6}$. For comparison,  the data for the solid state samples are presented. Here, we note that  the concentration is normalized by the surface area of the powder samples listed in Table \ref{T1}. 
The  CO$_{2}$ concentrations  for  both  the parent and $x=0.1$ citrate samples show a  rapid rise  at the initial 20 min under the visible light irradiation and then increase gradually with  increasing irradiation time.
The stable photocatalytic efficiency after 20 min  is related to the absorption process of  photogenerated CO$_{2}$   on the surface of  samples, which contributes to  reduction of CO$_{2}$  concentration.  
On the other hand,   for the Sb50$\%$ and   Sb100$\%$ substituted samples we detect no clear evolution.  
We suspect that the band gap opening due to  the heavy  Sb substitution  suppresses  the formation of electron-hole pairs, causing a decrease of  the photocatalytic reaction processes.   
In our previous study\cite{TA19},    the effect of the band gap opening due to the atomic substitution was examined by using first-principles electric structure calculation.   The Sb substitution at Bi site gives rise to a reduction of the Bi-orbitals and then results in the enlarged band gap.  The collapse of photocatalytic activity  is probably originated from the larger band gap, which is closely related to the removal of Bi orbitals in the electric structures.  The photocatalytic performance strongly depends on the Sb substitution, accompanied by an increase of the band gap energies. 
The photocatalytic data for the solid state samples with the same composition also show  similar behaviors as the citrate samples.  However,  the evolved   CO$_{2}$ values of the $x$=0 and $x$=0.1 citrate samples are more than twice  as  large as the data of the corresponding solid state samples,  as shown in  Fig.  \ref{CO2} ($b$).   The enhanced   photocatalytic properties  in the citrate samples are attributed to their morphology,  where  fine particles are  homogeneously distributed with a sub micron order.  
From the data listed in Table \ref{T1},   we conclude that the parent and low Sb substituted samples exhibit good catalytic performance among the same synthesis based material members. 

Finally, we discuss the  charge separation model for higher activities of the parent Ba$_{2}$PrBiO$_{6}$ compound,  as shown in Fig.\ref{Model}. 
As far as we know,  there are two approaches such as modification of  the band gaps and improvement of  the photogenerated charge separation  in the photocatalystic materials, for completing  high performance \cite{CH10,SU15,OH13,KA20}. 
We carried out the former  approach by larger Sb substitution at Bi site in the double perovskite parent compounds as mentioned before.  
Here,  we focus on  the valence mixing states between Pr$^{3+}$ and Pr$^{4+}$ from the  view point of charge separation.  
 As shown in  Fig.\ref{Model}, it is assumed that  the band gap at the cluster region occupied by the Pr$^{4+}$O$_{6}$ octahedra is narrower than that at its neighboring region occupied by the Pr$^{3+}$O$_{6}$ octahedra.   For  the Pr$^{4+}$ ions,   the valence of their surrounding Bi ions  is expected to change  from 5+ to  3+  under  electric neutral condition.  Accordingly, the band gap  becomes narrower   in the presence of the Pr$^{4+}$O$_{6}$ octahedra because  6s electrons are added to  empty  orbitals of Bi$^{5+}$  contributing to a majority of valence and conduction bands\cite{TA07}.    The ptotoexcited electrons from the lower valence to upper conduction band do not recombine with holes,  but transfer to its neighboring  lower conduction band, resulting in charge separation.  
It has been reported that the Pr based compound with different valence states  exhibits the highest efficiency among Ba$_{2}$ReBiO$_{6}$ series (Re=rare earth ions such as La, Pr, Ce, Nd, Sm, Eu, Gd, and  Dy) \cite{HA10}.   In previous studies on the crystal structure and magnetic properties of  Ba$_{2}$ReBiO$_{6}$,  the Pr ion only is shown to  exist in trivalent and tetravalent states (except for Ce)\cite{OT15}. 
Therefore, we believe that our  proposed model  qualitatively describes one of the possible  reasons for the efficient performance of the parent Ba$_{2}$PrBiO$_{6}$ compound.

\section{SUMMARY}
We investigated the crystal structures, magnetic, and  photocatalytic properties of the B-site substituted double perovskite Ba$_{2}$Pr(Bi$_{1-x}$Sb$_{x}$)O$_{6}$ ($x$=0, 0.1, 0.2, 0.5 and 1.0) synthesized  by the  citrate pyrolysis precursor method.  For comparison,  the photocatalytic data for the solid-state sample were examined.  The present synthesis procedure enabled to prepare highly homogeneous and fine powders.  
The single-phase poly crystalline samples with  the light Sb substitution crystallized  in  a monoclinic structure ($I2/m$). 
Magnetization measurements showed that the effective magnetic moments are located  around 3  $\mu_{B}$ , indicating the valence mixing states between Pr$^{3+}$ and Pr$^{4+}$. 

 The magnitudes of band gap energy for the two end member samples  were estimated from the optical measurements to be $E_{g}$ =1.06 eV at $x$=0  and  2.71 eV at $x$=1.0. The magnitude of the energy band gap is substantially enhanced with increasing  Sb content. When the Sb content exceeds 50 $\%$, the nature of photon absorption is changed from the indirect to direct transitions.
 The photocatalytic performance strongly depends on the Sb substitution accompanied by  an increase of the band gap energies. 
 The photocatalytic activities  of  the citrate  powder samples with the light Sb substitution are considerably   enhanced  in comparison with  those values of the samples prepared by  the conventional solid state method.  We believe that our  charge separation model  qualitatively describes one of the possible  reasons for the efficient performance of the parent Ba$_{2}$PrBiO$_{6}$ compound in the presence of trivalent and tetravalent states of the Pr ions. 
The mixed valence state of Pr ions is not only related to B site partial disordering in the monoclinic phase, but also has close relationship with occurrence  of the cubic phase in the phase diagram of the double perovskite compounds Ba$_{2}$PrSbO$_{6}$. 

We tried the two approaches to  controlling the band gaps and realizing  the photogenerated charge separation  in the photocatalystic materials through  larger Sb substitution and the valence mixing of the Pr ions.  Our findings suggest that the higher photocatalytic activities  strongly depend  on the powder preparation method as well as the modification of band gaps and the photo-induced charge separation. 
\\
\section{Acknowledge}
This work was supported in part by MEXT Grands-in-Aid for Scientific Research (JPSJ KAKENHI Grants No. JP19K04995),   Iketani Science and Technology Foundation, and The  Mazda Foundation. 
We thank Mr. K. Sasaki and M. Ito for the SEM measurement

Appendix A. Supplemental data

Supplemental data to this article can be found online at https://   .



\begin{thebibliography}{30}
\bibitem{VA15} S. Vasala, and M. Karppinen, {\it Prog. Solid State Chem.}, vol. 43, pp. 1-31, 2015.
\bibitem{KO98} K. -I. Kobayashi, T. Kimura, H. Sawada, K. Terakura, and Y. Tokura, {\it Nature}, vol. 395, pp. 677, 1998. 
\bibitem{RA07} R. Ramesh, and N. A. Spaldin, {\it Nature Materials}, vol. 2, pp. 21, 2007.
\bibitem{FU72} A. Fujishima, and K. Honda, {\it Nature}, vol. 238, pp. 37, 1972.
\bibitem{EN03} H. W. Eng,  P. W. Barnes,  B. M. Auerand, and  P. M. Woodward, {\it J. Solid State Chem.}, vol.175,  pp. 94, 2003. 
\bibitem{HA10} T. Hatakeyama,  S. Takeda,  F. Ishikawa,  A. Ohmura, A. Nakayama, Y. Yamada, A. Matsushita, and J. Yea, {\it J. Cer. Soc. Jpn. }, vol. 118,  pp.91-95, 2010.  
\bibitem{MA12} A. Matsushita, T. Nakane, T. Naka, H. Isago, Y. Yamada, and Yuh Yamada, {\it Jpn. J. Appl. Phys.}, vol. 51, pp. 121802-1-5, 2012. 
\bibitem{HA95} William T. A. Harrison, K. P. Reis, A. J. Jacobson, L. F. Schneemeyer, and J. V. Waszczak, {\it Chem. Mater.}, vol. 7, pp. 2161-2167, 1995.

\bibitem{SH76} R. Shannon, {\it Acta. Crystallogr.}, vol. A32, pp. 751, 1976. 
\bibitem{CH90}L. A. Chick, L. R. Pederson, G. D. Maupin, J. L. Bates, L/ E. Thomas, G. J. Exarhos, {\it Materials Letters}, vol. 10, 6-12, 1990.

\bibitem{KO91}K. Koyama, A. Junod, T. Graf, G. Triscone, and J. Muller, {\it Physica C}, vol.185-189, 66-70, 1991.
\bibitem{HA06}M. Hagiwara, T. Shima, T. Sugano, K. Koyama, and M. Matsuura, {\it Physica C}, vol. 445-448, 111-114, 2006. 
\bibitem{CH10}X. Chen, S. Shen, L. Guo, and S. Mao, {\it Chem. Rev.} vol.110, 6503-6570, 2010.
\bibitem{SU15}D. Sudha, and P. Sivakumar, {\it Chemical Engineering and Processing}, vol. 97, 112–133, 2015.
\bibitem{OH13}B. Ohtani, {\it Catalysts}, vol.3, 942-953, 2013.
\bibitem{KA01}M. Kawachi, N. Sato, E. Suzuki, S. Ogawa, K. Noto, and M. Yoshizawa, {\it Physica C}, vol. 357-360, 1023-1026. 
\bibitem{ON18} K. Onodera, T. Kogawa, M. Matsukawa, H. Taniguchi, K. Nishidate, A. Matsushita, and M. Shimoda, {\it J. Phys. Conf. Ser.}, vol. 969, pp. 012122-1-6, 2018. 

\bibitem{TA07} J. W. Tang, Z. G. Zou, and J. H. Ye,{\it J. Phys. Chem. C}, vol. 111, pp. 12779-12785, 2007.
\bibitem{MU04} T. Murase, H. Irie, and K. Hashimoto, {\it J. Phys. Chem. B},  vol. 108, 15803-15807 (2004)


\bibitem{FU05} W.T. Fu, and D.J.W. IJdo, {\it J. Solid State Chem.}, vol. 178, pp. 2363-2367, 2005.
\bibitem{OT15} S. Otsuka, and Y. Hinatsu, {\it J. Solid State Chem.}, vol. 227, pp. 132-141, 2015.
\bibitem{TA19} H. Taniguchi,M. Matsukawa, K. Onodera, K. Nishidate, and A. Matsushita,  {\it IEEE Transactions on Magnetics}, 55(2), 2019.
\bibitem{KA20}C. Karthikeyan, Prabhakarn Arunachalam, K. Ramachandran,
Abdullah M. Al-Mayouf, S. Karuppuchamy, {\it J. of Alloys and Compounds}, 828, 154281, 2020.

\end{thebibliography}
\end{document}